\documentclass[twocolumn,showpacs,prl,amsmath,amssymb]{revtex4}
\usepackage{colordvi,epsfig,bm}
\newcommand{\cN}{{\cal N}}

\newcommand{\bbbZ}	{{\mathbb{Z}}} 			
\newcommand{\bbbone}	{{\mathrm{1\hspace{-1mm}I}}}	

\newcommand{\rd} {\mathrm d}
\newcommand{\re} {\mathrm e}
\newcommand{\ri} {\mathrm i}

\newcommand{\bra}{\langle}
\newcommand{\ket}{\rangle}
\newcommand{\nn}{\nonumber} 
\newcommand{\be}{\begin{equation}} 
\newcommand{\ee}{\end{equation}}  
\newcommand{\bea}{\begin{eqnarray}}
\newcommand{\eea}{\end{eqnarray}}
\begin{document}
\title{Spectra of Empirical Auto-Covariance Matrices}
\author{R. K\"uhn and P. Sollich}
\affiliation{Department of Mathematics, King's College London, UK}
\date{\today}

\begin{abstract}
We compute spectra of sample auto-covariance matrices of second order 
stationary stochastic processes. We look at a limit in which both the matrix 
dimension $N$ and the sample size $M$ used to define empirical averages diverge, 
with their ratio $\alpha=N/M$ kept fixed. We find a remarkable scaling
relation which expresses the spectral density $\rho(\lambda)$ of sample 
auto-covariance matrices for processes {\em with\/} dynamical correlations 
as a continuous superposition of appropriately rescaled copies of the spectral 
density $\rho^{(0)}_\alpha(\lambda)$ for a sequence of {\em uncorrelated}
random variables. The rescaling factors are given by the Fourier transform 
$\hat C(q)$ of the auto-covariance function of the stochastic process. We 
also obtain a closed-form approximation for the scaling function $\rho^{(0)}_\alpha
(\lambda)$. This depends on the shape parameter $\alpha$, but is otherwise 
universal: it is independent of the details of the underlying random variables, 
provided only they have finite variance. Our results are corroborated by 
numerical simulations using auto-regressive processes.
\end{abstract}


\maketitle
The present investigation concerns spectral properties of sample auto-covariance 
matrices derived from time series, and in particular the way they are affected 
by finite sample fluctuations. Having a theory that would quantify such effects
analytically would clearly be useful for e.g.\ the empirical analysis of stochastic 
processes. However, such theoretical understanding is at present almost entirely
lacking -- in marked contrast to the situation for the closely related problem of 
sample covariance matrices of a multi-dimensional data set estimated from finitely 
many independent measurements.

From an abstract point of view, both problem classes belong to 
random matrix theory \cite{Mehta, Akemann+11}. In the case of sample covariance 
matrices, the random matrix ensemble in question is the well known Wishart-Laguerre 
ensemble \cite{Wish28}, which has been widely studied for several decades, and for which 
numerous results are available. The spectral problem for example was solved in the 60s by 
Mar\v{c}enko and Pastur \cite{MarPas67}; typical fluctuations of the largest eigenvalue 
of Wishart matrices were shown \cite{Johnst01} to follow a Tracy-Widom distribution 
\cite{TracWid96}, and large deviation properties of both the largest \cite{Vivo+07} 
and smallest \cite{KaPe10} eigenvalue have recently been characterized. Numerous variants 
of the original Wishart-Laguerre ensemble have been studied in the literature over 
the years (e.g. \cite{Burd+05, AkeViv08, BieTh08, Rech+10}), and applications formulated
in a variety of fields, including multivariate statistics \cite{Muir82}, wireless 
communication \cite{TolVerd04} and the analysis of cross-correlations in financial data 
\cite{Laloux+99, Plerou+99}. For a more complete recent overview, we refer to 
\cite{Akemann+11}.

Due to the temporal structure of the underlying signals in the problem of sample 
auto-covariance matrices of time series, the ensemble of random  matrices describing
this problem is radically different from the Wishart-Laguerre ensemble, and {\em much\/} 
less is known about their spectral properties. The {\em existence\/} of the limiting 
spectral density of sample auto-covariance matrices of moving-average processes \cite{Ham94}
with i.i.d.\ driving (of both finite and infinite order) has in fact been established 
only very recently \cite{Basak+11}; the corresponding existence proof for the closely 
related problem of random Toeplitz matrices with i.i.d.\ entries is also only a few years 
old \cite{Bryc+06}. We are not aware of closed form expressions for limiting 
spectral densities for these cases -- whether exact, or approximate
but of a quality that would allow meaningful use for e.g.\ time series analysis. The purpose 
of the present letter is to report recent progress that fills this gap.

We consider stationary zero-mean processes $(x_n)_{n\in\bbbZ}$. These could be 
discrete-time processes to begin with, or sampled from continuous-time processes
at discrete equidistant time steps $\Delta \tau$, in which case 
$x_n\equiv x(n\Delta \tau)$. We are interested in the spectrum of $N\times N$ 
empirical auto-covariance matrices $C$, which are estimated by
measurements on sequences of $M$ samples. 
There are several (non-equivalent) ways to define the elements of $C$. 
Our choice is
\be
C_{k\ell}= \frac{1}{M} \sum_{m=0}^{M-1} x_{m+k} x_{m+\ell}\ , \quad 1\le k,\ell\le N\ .
\label{defC}
\ee
Sample auto-covariance matrices $C$ of this form constitute randomly perturbed Toeplitz 
matrices \cite{GreSze84}. Note that our choice differs from the ones in \cite{Basak+11}, 
which are simpler for being constructed as random Toeplitz matrices from the start.

Our main results are the following. We find a remarkable scaling relation which
expresses the spectral density $\rho(\lambda)$ of sample auto-covariance matrices for 
processes {\em with\/} dynamical correlations as a continuous superposition of 
rescaled copies of the spectral density $\rho^{(0)}_\alpha(\lambda)$ for a sequence of {\em 
uncorrelated}, i.i.d.\ random variables. We also obtain a simple closed form 
expression for $\rho^{(0)}_\alpha$ that provides an excellent
approximation to numerically simulated spectra.

The spectral density of a matrix $C$ is evaluated in terms of its resolvent as
\be
\rho_N(\lambda ; C)= \frac{1}{\pi N}{\rm Im ~ Tr} ~
\big[\lambda_{\varepsilon} \bbbone - C\big]^{-1}\ .
\label{specC}
\ee
Here $\bbbone$ is the $N\times N$ unit matrix and $\lambda_{\varepsilon} = 
\lambda - i\varepsilon$, the limit $\varepsilon  \to 0^+$ being understood. 
We follow Edwards and Jones \cite{EdwJon76} and express the trace of the 
resolvent in terms of a Gaussian integral as
\be
\rho_N(\lambda) = -\frac{2}{\pi}~\lim_{\varepsilon\to 0} {\rm Im}~
\frac{\partial}{\partial \lambda} ~\frac{1}{N}\left\langle \ln Z_N\right\rangle\ ,
\label{rhoN}
\ee
with 
\be
Z_N = \int \prod_{i=k}^N \frac{\rd u_k}{\sqrt{2\pi/\ri}}~\exp\Big\{
-\frac{\ri}{2}\sum_{k,\ell} u_k (\lambda_{\varepsilon} \delta_{k\ell} - 
C_{k\ell})u_\ell\Big\}
\label{Zn}
\ee
and angled brackets indicating an ensemble average. This average can be evaluated 
using replicas. Analogous calculations in random matrix theory \cite{EdwJon76} suggest
that the final results will exhibit the structure of a replica-symmetric high-temperature 
solution, and hence that an annealed calculation (which replaces $\bra 
\ln Z_N \ket$ by $\ln \bra Z_N\ket$ in (\ref{rhoN})) will provide exact results.
This is the approach we adopt here.

Inserting the definition Eq.~(\ref{defC}) into Eq.~(\ref{Zn}), one notes that $Z_N$ 
depends on the disorder, i.e.\ on the $\{x_n\}$, only through the $M$ variables
\be
z_{i}  = \frac{1}{\sqrt N}\sum_{k=1}^N x_{i+k}  u_{k}\ ,\quad 0\le i <M\ .
\ee
Assuming that the true auto-covariance $\bar C(k)=\bra x_n x_{n+k}\ket$ is absolutely
summable, we can argue from the central limit theorem (CLT) for weakly dependent random 
variables that the $z_i$ will be correlated Gaussian variables with $\bra z_{i}\ket = 0$, 
and covariance matrix $Q$ whose elements are given in terms of $\bar C$ as
\be
Q_{ij} = \bra z_{i}z_{j}\ket = \frac{1}{N} \sum_{k\ell}\,\bar
C(i-j+k-\ell)
\,u_{k}u_{\ell }\ .
\label{defQ}
\ee
The disorder average $\bra \dots\ket$ is thus a Gaussian integral, which can be 
performed to give
\bea
\bra Z_N\ket &=& \int \prod_{k}\frac{\rd u_{k}}{\sqrt{2\pi/\ri}}\
\exp\Big\{-\frac{\ri}{2}\lambda_\varepsilon \sum_{k} u_{k}^2 \nn\\
& & ~~~~~~~~~~~~~~~~~~ -\frac{1}{2} \ln {\rm det}(\bbbone - \ri\alpha Q)\Big\}\ .
\label{Zeval}
\eea

The matrix $Q$ being Toeplitz, we will use Szeg\"o's theorem \cite{GreSze84} to 
evaluate the `spectral sum' $\ln {\rm det} (\bbbone - \ri\alpha Q)$ in Eq.\ (\ref{Zeval}). 
Given that our sequence of $Q$-matrices doesn't fully fit the assumptions of the standard 
theory in that matrix elements are {\em themselves} dependent on 
the dimension $M$, we expect this to be only an approximation; it should, 
however, become exact in the limit $\alpha \to 0$.

To proceed, we need Fourier representations of $Q$, and we will have to keep keep track 
of finite-$M$, finite-$N$ expressions in what follows. Assuming $M$ to be odd, we 
have
\be
Q_{ij} =  \frac{1}{M}\sum_{\mu=-(M-1)/2}^{(M-1)/2} \re^{-\ri q_\mu(i-j)} 
Q_\mu 
\label{fourQ}
\ee
for the ($\{u_k\}$ dependent) elements of $Q$, with 
\bea
Q_\mu  &=& \frac{1}{N}\sum_{k\ell} \hat C(q_\mu) \re^{-\ri q_\mu\,(k-\ell)}
\,u_{k}u_{\ell}\nn\\
& = & \hat C(q_\mu) |\hat u(q_\mu)|^2 \equiv Q(q_\mu)
\eea
where $q_\mu=\frac{2\pi}{M}\,\mu$ and $\hat u(q_\mu) = \frac{1}{\sqrt N} 
\sum_{k=1}^N \re^{\ri q_\mu k} u_k$. Here
\be 
\hat C(q_\mu) = \sum_{n=- (M-1)/2}^{(M-1)/2} \bar C(n) \re^{\ri q_\mu n}
\label{fourC}
\ee
is the Fourier transform of the true auto-covariance function of the underlying process. 
Truncating the sum in Eq.\ (\ref{fourC}) at $|n| \le (M-1)/2$ will create negligible errors in 
the large $M$ limit if $\sum_{n=-\infty}^\infty |\bar C(n)|$ exists,
as already required when appealing to the  
CLT for the $z_i$ statistics above. Restricting the $q_\mu$ values to the discrete grid 
with spacing $2\pi/M$ approximates  $Q$ by its cyclified version. In Szeg\"o's terminology, 
the matrix $Q$ has $Q_\mu=Q(q_\mu)$ as its ($M$-grid) symbol, and Szeg\"o's approximation for 
the spectral sum reads
\be
\ln {\rm det} (\bbbone - \ri\alpha Q) \simeq \sum_{\mu=-(M-1)/2}^{(M-1)/2} \ln \big(1-\ri 
\alpha Q_\mu\big) \ .
\label{specsQ}
\ee
The symmetry $\bar C(n)=\bar C(-n)$  entails $\hat C(q_\mu)=\hat C(-q_\mu)$, thus $Q(q_\mu) 
=Q(-q_\mu)$. Next, one extracts the $\{u_k\}$ dependence (via the $\{Q_\mu\}$) from the 
evaluation of (\ref{specsQ}), using $\delta$-functions and their
Fourier representations.
The $\{u_k\}$ integrals then become Gaussian, and we can express $\bra Z_N \ket$ as
\bea 
\bra Z_N \ket &=& \int\prod_{\mu=0}^{(M-1)/2}\frac{\rd\hat Q_\mu \rd Q_\mu}{2\pi}\ 
\exp\Big\{\!-\! \sum_{\mu=0}^{(M-1)/2} \ri \hat Q_\mu Q_\mu \nn\\
& &\hspace{-9mm} -\!\! \sum_{\mu=0}^{(M-1)/2} \ln (1 - \ri\alpha Q_\mu)-\frac{1}{2} 
\ln {\rm det} (\lambda_\varepsilon \bbbone - R) \Big\}.
\label{ZNR1}
\eea 
The elements of the $N\times N$ matrix $R$ in (\ref{ZNR1}) are
given by $R_{k\ell} = \frac{2}{N}\sum_{\mu=0}^{(M-1)/2} \hat Q_\mu \hat C(q_\mu) 
\cos(q_\mu(k-\ell))$, with $1\le k,\ell\le N$. We have combined modes with
$\mu$ and $-\mu$ and neglected as subleading the fact that the
resulting prefactors differ for the $\mu=0$ mode.

\begin{figure}[t]
\epsfig{file=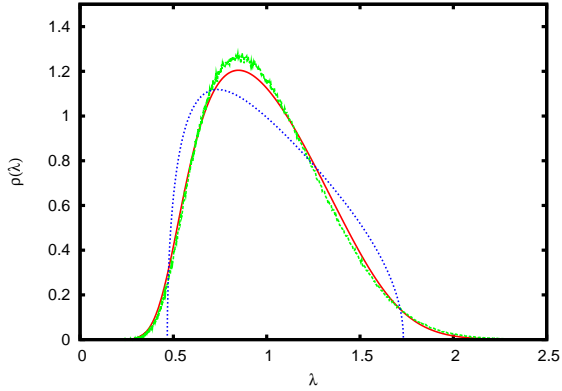, width=0.45\textwidth}
\caption{(Colour online) Spectral density for sample auto covariance matrices of i.i.d.\ 
signals $x_n \sim \cN(0,1)$ at $\alpha =0.1$ (green); analytic approximation Eq.\ (\ref{rho0}) 
for $\rho_\alpha^{(0)}(\lambda)$ (red).  The Mar\v{c}enko-Pastur law 
(blue-dashed) for the same $\alpha$ is also shown for comparison.}
\end{figure}

We use residues to evaluate the $Q_\mu$ integrals in (\ref{ZNR1}):
$$
\int \frac{\rd Q_\mu}{2\pi}\, \frac{\re^{-\ri \hat Q_\mu Q_\mu}}
{1-\ri \alpha Q_\mu} = \left\{\begin{array}{ll} 
\alpha^{-1}\, \re^{-\hat Q_\mu/\alpha } & 
;~~\hat Q_\mu >0\ ,\\ 0 & ;~~{\rm else}\ . \end{array}\right.
$$
After rescaling $\hat Q_\mu/\alpha \to \hat Q_\mu$ this yields
\bea 
\bra Z_N \ket &=& \left\langle \exp\Big\{ -{\textstyle\frac{1}{2}} \ln {\rm det} 
(\lambda_\varepsilon \bbbone - R) \Big\}\right\rangle_{\{\hat Q_\mu\}}
\label{ZNR}
\eea 
with now 
\be
R_{k\ell} = \frac{2}{M}\sum_{\mu=0}^{(M-1)/2} \hat Q_\mu \hat C(q_\mu)
\cos(q_\mu(k-\ell))\ .
\label{defR}
\ee 
In Eq. (\ref{ZNR}) we have introduced the short-hand
\be
\langle \dots \rangle_{\{\hat Q_\mu\}}= \int_0^\infty \prod_{\mu=0}^{(M-1)/2} 
\Big\{\rd\hat Q_\mu \re^{- \hat Q_\mu}\Big\}\, \big( \dots \big)
\ee
for the $\hat Q_\mu$-integrals. As the notation indicates, these amount to averages 
over exponentially distributed random variables of unit mean. Hence within our 
Szeg\"o-approximation, the original spectral problem for sample auto-covariance matrices 
$C$ is equivalent to that for random Toeplitz matrices $R$ given by (\ref{defR}).

To make progress we use the fact that the matrices $R$, too, are Toeplitz, and approximate 
the spectral sum $\ln {\rm det} (\lambda_\varepsilon \bbbone - R)$ appearing in (\ref{ZNR}) 
in terms of Szeg\"o's theorem,
\be
\ln {\rm det} (\lambda_\varepsilon \bbbone - R) \simeq 
\sum_{\nu =-(N-1)/2}^{(N-1)/2}\ln \big(\lambda_\varepsilon - R_\nu\big)\ ,
\label{specsR}
\ee
with 
\bea
R_\nu \! &=& \!\frac{1}{M}\sum_{\mu=0}^{(M-1)/2} \hat Q_\mu \hat C(q_\mu) \!\!\!\!\!
\sum_{n=-(N-1)/2;\sigma=\pm 1}^{(N-1)/2} \re^{\ri(p_\nu+\sigma q_\mu) n}
\nn\\
&=&\sum_{\mu=0}^{(M-1)/2} \hat Q(q_\mu) \hat C(q_\mu) S_{\nu\mu} \equiv R(p_\nu)\ ,
\label{Rnu}
\eea
for $p_\nu=\frac{2\pi}{N}\, \nu$ defined on a grid of spacing $2\pi/N$, and the $S$-kernel
given by
\be
S_{\nu\mu} = \frac{1}{M}\sum_{\sigma=\pm 1}\frac{\sin(N(p_\nu+ \sigma q_\mu)/2)}
{\sin((p_\nu+ \sigma q_\mu)/2)}\ .
\ee
Next one extracts the $\hat Q_\mu$ dependence from the spectral sum (\ref{specsR}) by enforcing
the $R_\nu$-definitions using $\delta$-functions and their Fourier
representations. This enables one to 
perform the $R_\nu$ integrals using residues much as in the case of the $Q_\mu$ integrals above, giving
\be
\bra Z_N \ket = \left\langle \prod_{\nu=0}^{(N-1)/2} F_\nu\right\rangle_{\{\hat Q_\mu\}}
\ee
with
\be
F_\nu = \ri \int_0^\infty \!\!\!\rd\hat R_\nu\, \re^{-\ri \hat R_\nu \big(\lambda_\varepsilon -
\sum_{\mu=0}^{(M-1)/2} \hat Q_\mu\hat C(q_\mu)S_{\nu\mu}\big)} \ .
\label{defFnu}
\ee
The coupling via the $S$-kernel entails that the $F_\nu$ for different $\nu$ are correlated. 
To proceed, we exploit the property that the $S$-kernel is rapidly oscillating, and sharply peaked 
at $|p_\nu \pm  q_\mu|\simeq {\cal O}(1/N)$. The dominant contributions to the exponential 
in (\ref{defFnu}) at fixed $\nu$ therefore lie in the interval $I_\nu=\{\mu: |\nu \pm \alpha \mu| \le 
1\}$. Approximating the $S$-kernel by a rectangular window (of height $\alpha/2$) on $I_\nu$
and using smoothness of $\hat C(q_\mu)$ on the $q_\mu$-scale, we set
\be
\sum_{\mu=0}^{(M-1)/2} \hat Q_\mu \hat C(q_\mu) S_{\nu\mu} \simeq \frac{\alpha}{2} \hat C(p_\nu) 
\sum_{\mu \in I_\nu} \hat Q_\mu\ .
\label{ovbin}
\ee
As the $I_\nu$ are overlapping, the $F_\nu$ in (\ref{defFnu}) remain correlated. As a last 
step we ignore these residual correlations and substitute $y=\alpha\hat R_\nu \hat C(q_\mu)/2$ 
in Eq.\ (\ref{defFnu}) to arrive at a closed form approximation for $\bra Z_N \ket$:
\be
\bra Z_N \ket \simeq \!\!\! \prod_{\nu=0}^{(N-1)/2}\!\!\!  \Bigg\{\frac{2\,\ri}{\alpha 
\hat C(p_\nu)} \int_0^\infty \!\!\!\!\!  \rd y \, \frac{\re^{-\ri y \lambda_\varepsilon 
2/(\alpha\hat C(p_\nu))}}{\big(1-\ri y\big)^{2/\alpha}} \Bigg\}
\label{Zfin}
\ee

\begin{figure}[t!]
\epsfig{file=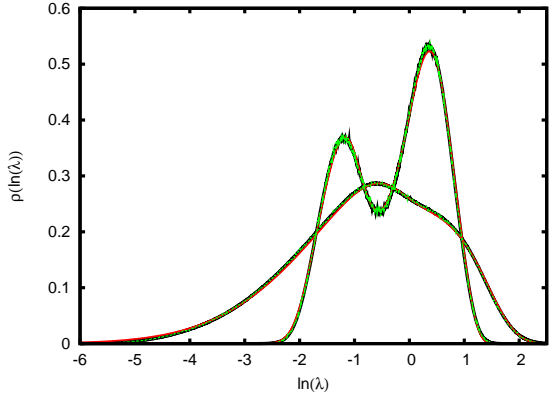, width=0.45\textwidth}
\caption{(Colour online) Logarithmic spectral densities of auto-covariance matrices for an
AR2 process  with $a_1=1/2$ and $a_2=5/16$, comparing the scaling prediction Eq.\ (\ref{rhofin}) 
using the empirical scaling function (see text, black full curve) with that based on the analytic 
approximation (\ref{rho0}) for the scaling function (red full curve), and simulations (green  
dashed curve) Double-peaked set of curves: $\alpha=0.1$, single-peaked set of curves: $\alpha=0.8$.}
\end{figure}

For the spectral density (\ref{rhoN}) in the thermodynamic limit $N\to \infty$ we then get
\bea
\rho(\lambda) &=& -\frac{2}{\pi} \lim_{\varepsilon\to 0}\, {\rm Im} \frac{\partial}{\partial 
\lambda}  \lim_{N\to\infty} \frac{1}{N} \ln \bra Z_N \ket\nn\\
&=& \int_0^\pi \frac{\rd q}{\pi}\, \frac{1}{\hat C(q)}\,\rho_\alpha^{(0)}(\lambda/\hat C(q))
\label{rhofin}
\eea
in which 
\be
\rho_\alpha^{(0)}(\lambda) = - \lim_{\varepsilon\to 0}\,\frac{1}{\pi} {\rm Im} \frac{\partial}
{\partial \lambda} \ln  I_\alpha\bigg( \frac{2}{\alpha}\lambda_\varepsilon\bigg)
\label{rho0}
\ee
with $I_\alpha$ obtained from (\ref{Zfin}) in terms of an incomplete $\Gamma$-function: for 
${\rm Im}\, x <0$,
\bea
I_\alpha(x) &\equiv& \int_0^\infty \rd y \, \re^{-\ri y x}\big(1-\ri y\big)^{-2/\alpha}
\nn\\
& = &\ri \, (-x)^{-1+2/\alpha}\, \re^{-x} \Gamma(1-2/\alpha,-x)\ .
\label{defIal}
\eea
Note that the scaling function $\rho_\alpha^{(0)}(\lambda)$ has an
independent meaning as 
the spectral density of the empirical auto-covariance matrix (at the same value of the shape parameter 
$\alpha$) for a sequence of {\em uncorrelated\/} data, for which $\hat
C(q)\equiv 1$. Eq.\ (\ref{rhofin}) thus constitutes a remarkable scaling relation 
relating the spectral density of sample auto-covariance matrices for processes with 
dynamical correlation to the spectral density of sample auto-covariance 
matrices for i.i.d.\ sequences of random data.

We have checked our results using simulations of sets of 5000 $N\times N$ auto-covariance matrices 
with $N=800$.  Fig.\ 1 compares simulations for a sequence of i.i.d.\ variables with our prediction 
(\ref{rho0}) for $\rho_\alpha^{(0)}$, and the Mar\v{c}enko-Pastur law at $\alpha=0.1$. Fig.\ 2 looks 
at  auto-regressive AR2 processes of the form $x_n + a_1 x_{n-1}+a_2 x_{n-2} = \sigma \xi_n$, with 
i.i.d.\ $\xi_n \sim \cN(0,1)$, and $\sigma$ chosen to ensure that $\bar C(0)=\bra x_n^2 \ket =1$. It
compares simulations with scaling based on our analytic approximation (\ref{rho0}) for $\rho_\alpha^{(0)}$, 
and with scaling using an empirical scaling function determined via simulation, with an $\alpha=0.1$-example 
shown in Fig.\ 1. 

Whereas scaling appears to be exact using the empirical scaling function, our analytic result is
not. It nevertheless produces quantitatively accurate results, even for $\alpha$ as large as 0.8. 
We have elements of an independent proof of scaling which we intend to publish in a forthcoming paper. 
Judging from the impact that analogous results for (Wishart-Laguerre) sample covariance matrices have had,
we believe our results to hold significant potential for applications in a variety of fields, including 
time-series analysis, information theory, or signal processing.

{\bf Acknowledgement:}  It is a pleasure to thank K.\ Anand, L.\ Dall'Asta and P.\ Vivo 
for illuminating discussions on the occasion of a visit of RK to the ICTP at Trieste, 
which triggered the present investigation.

\bibliography{../../MyBib}
\end{document}